# From the First to Subsequent Pulses: Evolution of Discharge inside a Preformed Bubble in Water

Yang Xia[1], Siyuan Liu[1], Zhanqiang Liu[1], Weishan Zhang[1], Zhihua Qi[2]

[1] School of Physics and Electronic Technology, Liaoning Normal University, Dalian, People's Republic of China 116029
[2] Basic Teaching Department, Shenyang Institute of Engineering, Shenyang, 110136, China

## Abstract

The evolution of pulsed discharge behavior inside a preformed air bubble in water from the first to subsequent pulses was experimentally investigated using a synchronized needle–bubble system. A positive nanosecond high-voltage pulsed power supply, together with a pulse valve and ICCD imaging, was employed to generate reproducible preformed bubbles and to record the corresponding discharge development with good temporal synchronization. The results show that, although the preformed bubbles exhibit good repeatability in size and morphology under identical conditions, the first-pulse discharge inside the bubble remains highly stochastic. The first discharge is predominantly corona-like and is not significantly affected by bubble size once the electrode is covered by the bubble. By varying the pulse width, the discharge inside the bubble was observed to evolve progressively from corona-like emission to streamer discharge, accompanied by increasing instability of the bubble interface. At sufficiently large pulse width and pulse number, bubble wrinkling and even rupture were induced. The effect of solution conductivity was also examined. Increasing conductivity significantly enhanced discharge intensity, enlarged the luminous region, and promoted streamer propagation along the inner bubble surface. At sufficiently high conductivity, the first pulse already produced strong discharge and rapid bubble rupture. In addition, the current amplitude and the energy dissipated per pulse increased with conductivity and pulse number. These results demonstrate that the discharge evolution inside a preformed bubble is jointly governed by pulse history, pulse width, and solution conductivity, and that residual effects from previous pulses play an important role in the transition from the first pulse to subsequent discharges.

# 1.Introduction

Electrical discharges in water and at water–gas interfaces have attracted sustained attention in atmospheric-pressure plasma physics because they enable the simultaneous generation of strong electric fields, ultraviolet radiation, shock waves, and chemically reactive species in a compact system [1,2]. Such discharges are relevant to water treatment, sterilization, advanced oxidation, plasma-assisted chemistry, and materials processing[3-6]. Compared with conventional gas discharges, however, discharge initiation and propagation in water are considerably more complex because water is a dense polar medium with high collision frequency, strong electron energy loss, and strong coupling among plasma formation, heat transfer, phase change, and hydrodynamics [1,2,7]. As a result, the fundamental physics of water discharge remains an active topic, especially under pulsed excitation where highly transient, non-equilibrium conditions can be established before full thermalization occurs.

A central issue in underwater discharge research is the mechanism of breakdown initiation. In many practical systems, the discharge does not develop in a completely homogeneous liquid phase, such as dissolved gas, microcavities, vapor layers, or bubbles generated by electrolysis and Joule heating[8-10]. Reviews by Sunka , Akiyama, and Kolb et al. have emphasized that cavity-mediated or bubble-assisted mechanisms are crucial for interpreting a large fraction of pulsed discharges in water[7,11,12]. This is physically reasonable because a gaseous region provides a much more favorable environment for electron multiplication than bulk liquid water. Consequently, one effective strategy for studying underwater plasma is to deliberately introduce a preformed air bubble and ignite the discharge with a needle electrode inside or adjacent to that bubble. In this configuration, the initial gas phase is better controlled than in self-generated vapor channels, making it easier to isolate the role of the gas cavity in the onset and development of the discharge.

The needle–bubble configuration is particularly valuable because it combines strong field enhancement at the needle tip with a predefined low-density discharge region. The strong curvature of the needle electrode concentrates the electric field at the tip, while the gas bubble lowers the effective breakdown threshold relative to direct liquid-phase breakdown [13,14]. In addition, the large permittivity contrast between gas and liquid redistributes the local electric field near the bubble boundary, so the gas–liquid interface often becomes an important region for discharge initiation and propagation [15]. Under pulsed voltage, discharge inception in such systems usually begins near the needle tip, followed by streamer-like or corona-like emission inside the bubble and/or along the bubble interface. As the deposited energy increases, the discharge may evolve into a brighter filamentary channel, accompanied by bubble deformation, local vaporization, and eventual bridging of the inter electrode gap [17,16]. Therefore, in bubble-assisted underwater discharge, electrical evolution, optical evolution, and bubble dynamics are intrinsically coupled.This coupling also implies that the discharge state is not fixed, but may evolve through several regimes depending on the applied voltage, pulse width, rise time, repetition frequency, polarity, and electrode geometry [12-14]. In general, underwater pulsed discharges may proceed from a non-discharge state to intermittent inception, then to a streamer-like state, and finally to a highly conductive spark-like or arc-like state if energy deposition becomes sufficiently large [12-14]. In the context of a preformed air bubble, this state evolution is especially important because the bubble provides an initial gaseous path, but the subsequent plasma behavior still depends on whether the discharge remains confined inside the bubble, propagates along the interface, or drives further vapor expansion into the surrounding liquid. Thus, studying the evolution of discharge in a preformed air bubble is not only useful for understanding plasma formation itself, but also for clarifying how gas-phase ionization, interfacial physics, and liquid-phase energy deposition interact during pulsed breakdown.

Among the external parameters governing this process, solution conductivity plays a first-order role. Conductivity determines the liquid resistance, the conductive current, the RC charging time of the discharge gap, and the rate of ohmic heating in the liquid [11,17]. For this reason, conductivity affects both the inception stage and the later transition between discharge states. Importantly, the effect of conductivity is not universally monotonic. In a preformed bubble reactor, conductivity further affects how the bubble couples electrically to the surrounding liquid, making it a key parameter for interpreting both the discharge onset and the subsequent discharge-state transition.

Although substantial progress has been made in plasma–liquid interaction research, the pulse-to-pulse evolution of discharge behavior inside a preformed bubble remains insufficiently understood. Most previous studies have emphasized direct liquid breakdown, reactor performance, or time-averaged chemical outcomes, whereas fewer have examined the discharge from the initial pulse to later pulses in a controlled needle–bubble configuration. Therefore, a systematic investigation of underwater pulsed discharge in a needle–bubble system is needed to clarify how a preformed bubble governs discharge inception in the first pulse, how the discharge behavior develops over subsequent pulses. Such understanding is important for both the fundamental physics of repetitive plasma generation in gas–liquid systems and the optimization of plasma reactors operated in or near water.

## 2. Experimental setup

Figure 1 shows the experimental setup and the synchronized timing scheme used in this study. As shown in Figure 1(a), the system consists of a positive nanosecond high-voltage(H.V.) pulsed power supply, a pulse valve(Key high vacuum products, PEV-1), an ICCD camera (Andor, istar DH712), an oscilloscope (Tektronix, DPO5104), and a digital delay generator (Stanford Research Systems, DG645). The H.V. pulsed power supply is composed of a signal generator (Tektronix, AFG3022C), a DC high-voltage(H.V.) power supply (Spellman, SL10PN2000), and a pulse generator (DEI PVX-4110). A monochromatic sodium lamp is used as the backlight source to clearly visualize the bubble morphology and its relation to discharge behavior. The discharge electrode is a 500 μm tungsten wire inserted into a quartz tube with an inner diameter of 2.0 mm and an outer diameter of 4.0 mm, with the wire extending about 500 μm beyond the tube outlet. The experiments are conducted in a grounded aqueous medium. Deionized water is used as the base liquid, and its conductivity is adjusted in subsequent experiments by adding KCl.

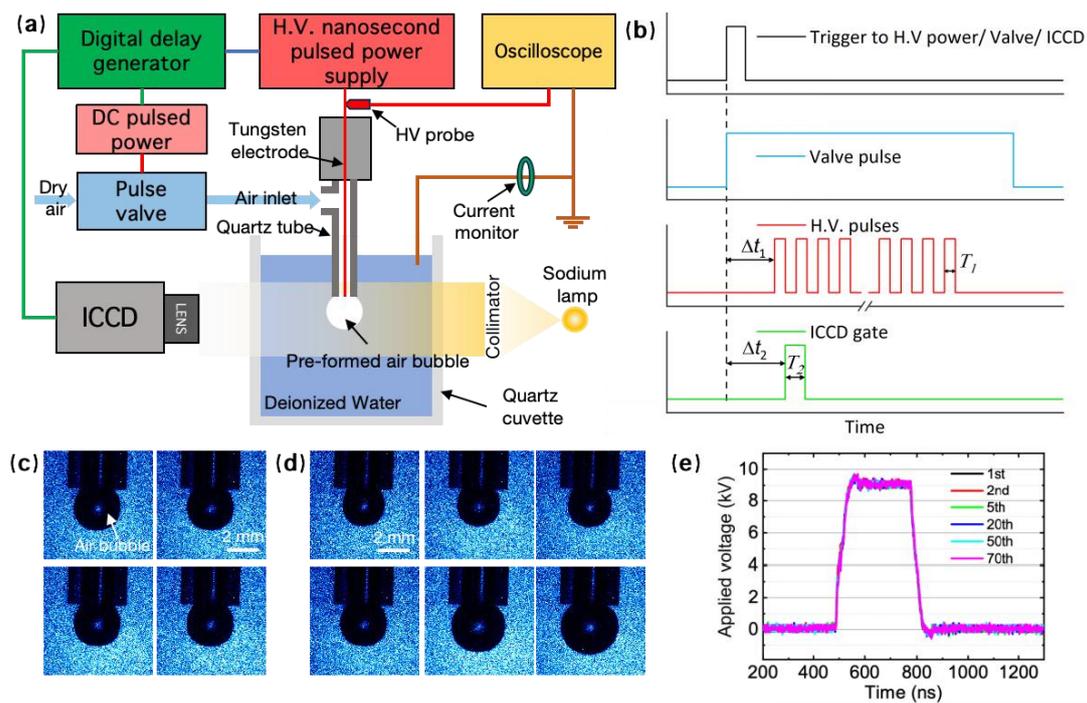

Figure 1. (a) Schematic of the experimental setup. (b) Schematic timing diagram of the experimental trigger and pulse sequence. (c) Preformed air bubbles generated at a fixed delay time, showing good repeatability. (d) Morphology of preformed air bubbles at different pulse numbers. (e) Typical applied-voltage waveforms for different pulse numbers at a pulse width of 300 ns.

Figure 1(b) shows the schematic timing diagram of the experimental trigger and pulse sequence. Trigger signals generated by the digital delay generator are used to synchronize the H.V. power supply, the pulse valve, and the ICCD camera. By controlling the valve pulse, a preformed air bubble is generated at the outlet of the quartz tube. After a delay of $\Delta t_1$ set by the H.V. power, a train of H.V. pulses is applied during the valve-open period, so that the discharge can be generated in the preformed air bubble at the tube outlet. The ICCD gate is opened after a delay of $\Delta t_2$ with a gate width of $T_2$, enabling time-resolved image acquisition of the bubble morphology and discharge development. As shown in Figure 1(c), preformed air bubbles generated at a fixed delay time exhibit good repeatability. The selected images indicate that the bubble size and shape are highly consistent under the same operating conditions, confirming the stability of the synchronized bubble-generation method. Figure 1(d) presents the morphology of preformed air bubbles at different pulse numbers within the voltage pulse sequence. The bubble size gradually increases with increasing pulse number while maintaining a relatively rounded shape, indicating continuous bubble growth during the pulse sequence. Figure 1(e) shows the typical applied-voltage waveforms for different pulse numbers at a pulse width ($T_1$) of 900 ns. The waveforms are generally similar for different pulse numbers, providing a consistent electrical excitation condition for investigating the evolution of bubble morphology and discharge behavior.

## 3. Results and discussion

Figure 2 shows representative discharge images inside the bubble and the discharge probability under different applied voltages. As shown in Figure 2(a), although the preformed bubbles are highly reproducible in both size and morphology under identical experimental conditions, shown in Figure 1(c), the specific discharge initiation location and the branching path of the streamer channels inside the bubble still exhibit strong spatial randomness from shot to shot. The discharge mainly propagates through the interior of the bubble in the form of curved or branched plasma channels. Meanwhile, due to the randomness of the discharge path and energy deposition inside the bubble, the bubble size after discharge is not identical from shot to shot, and slight variations in bubble morphology can be observed even under the same operating conditions. The initiation of an electrical discharge within a gas bubble submerged in a liquid is fundamentally governed by the stochastic availability of seed electrons, which are required to trigger initial electron avalanches[18]. Unlike discharge in an open gas gap, an isolated bubble is encapsulated by an insulating dielectric liquid, which prevents the continuous injection of free charges from the external environment. Consequently, a possible source of these seed electrons is random ionizing radiation from cosmic rays and natural background radioactivity.

Figure 2(b) shows the discharge probability as a function of the serial number of the HV pulse at different voltages. The discharge probability was determined optically using the ICCD camera. For each discharge pulse, optical signals were recorded 200 times, and the discharge was considered successful when a distinct luminous signal was observed in the ICCD image. The discharge probability was then calculated as the ratio of successful discharges to the total number of measurements. The discharge probability increases with both the applied voltage and the pulse number. At relatively low voltage, such as 3.0 kV, the discharge probability rises gradually with pulse number. In contrast, at higher voltages, the discharge probability increases much more rapidly and approaches 100% within fewer pulses, indicating that increasing the applied voltage significantly promotes discharge inception inside the bubble. The increase in discharge probability with pulse number is likely related to the accumulation of residual charges and metastable species during consecutive pulsing, which can significantly reduce the influence of the intrinsic randomness of breakdown initiation on the discharge process.

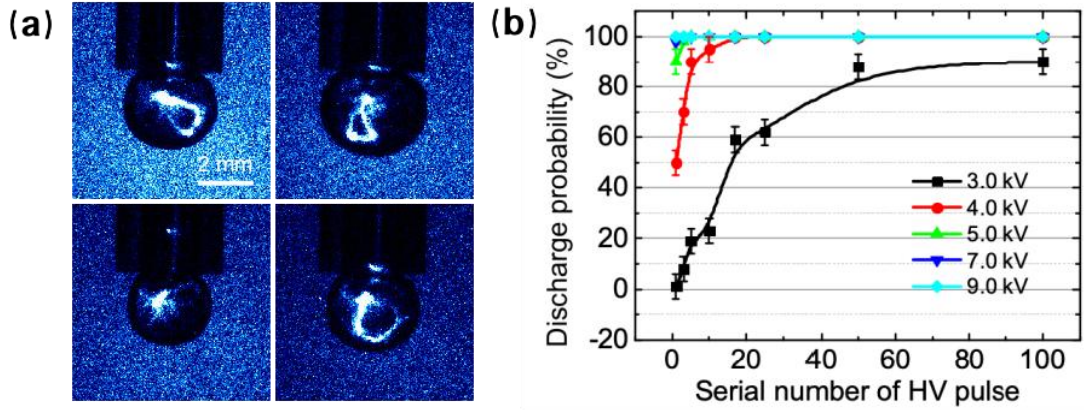

Figure 2. (a) Bubble discharge images under identical conditions, showing random discharge paths ($V_{amp}$ = 9.0 kV, $T_1$ = 2.0 μs, single shot, gate width = 5 μs). (b) Discharge probability versus HV pulse number at different voltages.

Figure 3 shows the first-pulse discharge images obtained in six preformed bubbles with different sizes in deionized water. Although the bubble size increases continuously across the six cases, no obvious change is observed in the emission intensity of the first-pulse discharge. When the preformed bubble does not fully cover the electrode, the discharge is produced inside the bubble in a corona-like mode. After the protruded bubble completely covers the electrode, the first discharge is still dominated by corona discharge, but it becomes mainly localized near the tip of the H.V. electrode. These results indicate that the first-pulse discharge is not significantly affected by bubble size.

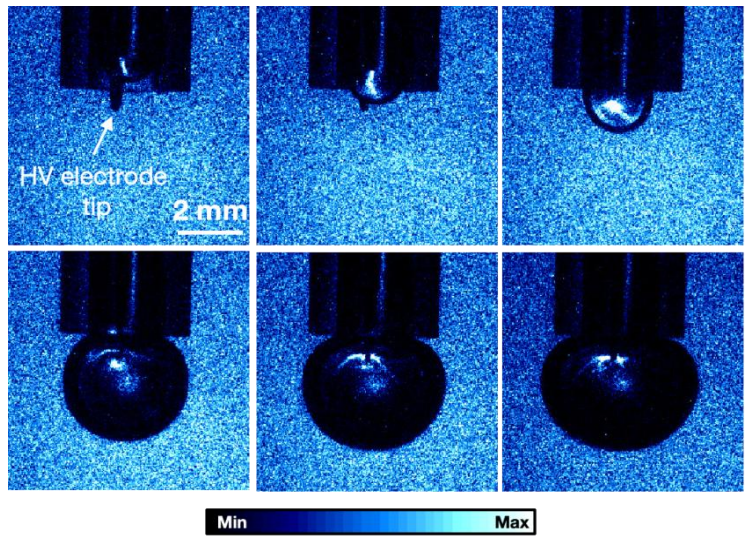

Figure 3. First-pulse discharge morphology in different-sized preformed air bubbles($V_{amp}$ = 9.0 kV, $T_1$ = 900 ns, single shot, gate width = 5 μs), obtained by adjusting $\Delta t_1$ and recorded by setting $\Delta t_2$.

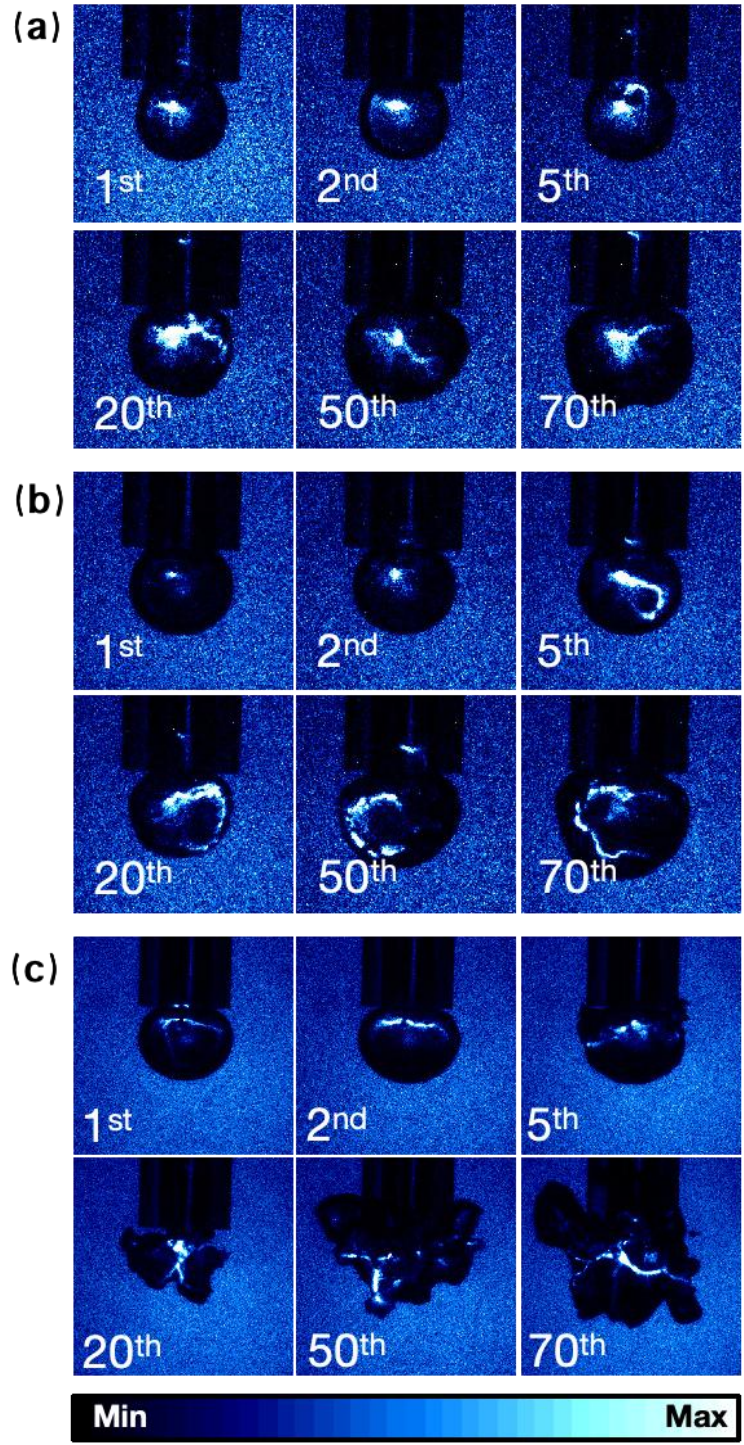

Figure 4. ICCD images of bubble discharge recorded at different pulse numbers under different pulse widths in deionized water with a conductivity of 4.6 μS/cm: (a) $T_1$ = 900 ns, gate width = 900 ns; (b) $T_1$ = 2.0 μs, gate width = 2 μs; and (c) $T_1$ = 20 μs, gate width = 20 μs. The images correspond to the 1st, 2nd, 5th, 20th, 50th, and 70th pulses, respectively.

To clarify the influence of residual seed charges on discharge development, the evolution of streamer discharge inside a bubble was investigated by varying the pulse width and pulse number of the applied voltage. Figure 4 shows the ICCD images of discharge inside a bubble in deionized water under different pulse widths and pulse numbers. When the pulse width is 900 ns, the discharge evolves from a corona mode to a streamer mode between the 1st and the 5th pulses. After more than 20 pulses, the bubble surface is no longer smooth and noticeable wrinkles appear near its edge, as shown in Figure 4(a). When the pulse width is increased to 2 μs, the discharge inside the bubble still gradually transforms from an initial corona mode into a non-uniform streamer mode, as shown in Figure 4(b). With a further increase in pulse width to 20 μs, the intensity and instability of the streamer discharge increase significantly with pulse number. When the discharge reaches the 20th pulse, the strong streamer channel formed inside the bubble induces bubble rupture, as shown in Figure 4(c).The wrinkling and rupture of a submerged bubble under pulsed excitation can be attributed to the coupled effects of thermal instability, surface charge accumulation, Maxwell stress, and discharge energy localization. Initially, the relatively low gas temperature and nearly uniform thermal field inside the bubble favor corona-like discharge. With increasing pulse width or pulse number, cumulative energy deposition increases the local gas temperature and decreases the gas density, which enhances the reduced electric field and promotes the transition from a relatively uniform corona to a strong non-uniform streamer discharge. Meanwhile, the deposition of charges on the gas–liquid interface produces Maxwell stress, which may destabilize the interface once it becomes comparable to the restoring force of surface tension, thereby inducing surface wrinkling[19]. As the discharge further develops, energy tends to concentrate in the strongest and earliest-formed streamer channels, which can propagate toward and even penetrate the interface into the surrounding liquid. Together with the pressure impulse, localized heating, and possible capillary-wave excitation caused by repetitive streamer impact, this process eventually leads to bubble rupture and fragmentation.

Figure 5 shows the variation of the maximum optical emission intensity with the serial number of high-voltage pulses for pulse widths of 300 ns, 900 ns, 2 μs, and 20 μs. For 300 ns and 900 ns, the first pulse produces a markedly higher maximum emission intensity than the subsequent pulses, after which the intensity rapidly decreases and remains nearly constant. In contrast, for 2 μs, the maximum intensity shows only weak variation over the pulse sequence. For 20 μs, the intensity exhibits larger fluctuations at low pulse numbers and an increasing trend as the pulse sequence proceeds.The larger standard deviations observed at 20 μs are attributed to discharge instability induced by bubble rupture, as shown in figure 4(c), which causes abrupt variations in the gas–liquid interface and consequently leads to strong pulse-to-pulse fluctuations in the maximum emission intensity.

The pronounced first-pulse enhancement at short pulse widths can be attributed to the fact that the initial discharge occurs in an undisturbed medium with maximum local field enhancement, resulting in a strong transient ionization event. After the first pulse, residual charges, local heating, and bubble interface evolution modify the local discharge conditions and reduce the peak emission intensity of the following pulses. As the pulse width increases, the discharge development becomes less controlled by the initial transient process and more affected by cumulative energy deposition and

gas–liquid interface evolution. Consequently, for the 20 μs case, the later pulses can sustain stronger emission, giving rise to the observed gradual increase in maximum intensity.

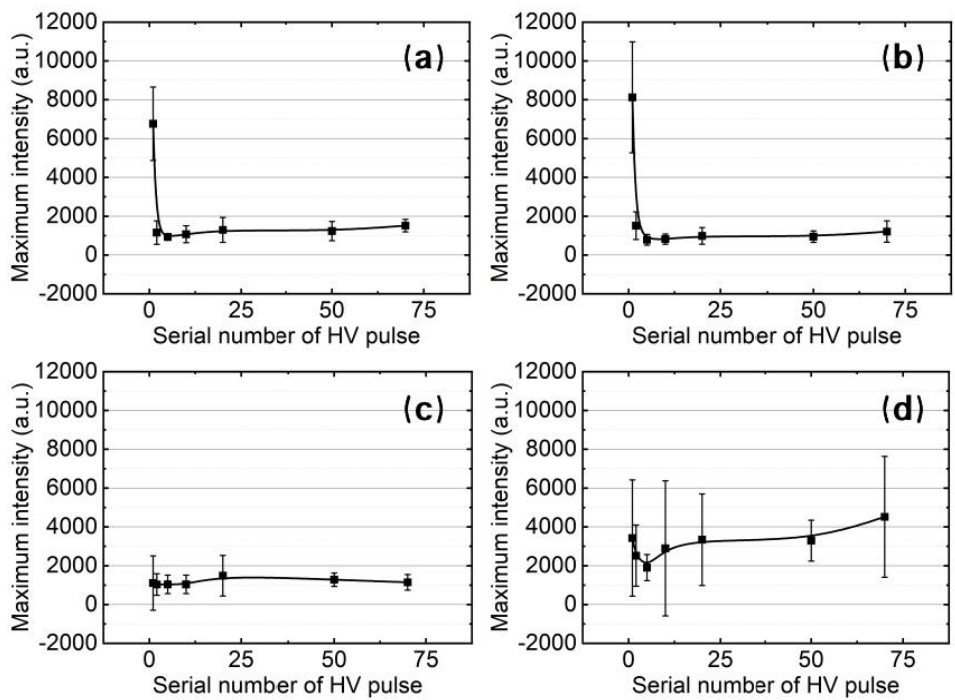

Figure 5. Variation of the maximum optical emission intensity with the serial number of high-voltage pulses under repetitive-pulse discharge conditions for different pulse widths: (a) 300 ns, (b) 900 ns, (c) 2 μs, and (d) 20 μs. The data points represent the mean maximum emission intensity, and the error bars indicate the standard deviation.

In practical discharge processes, the solution conductivity does not remain constant but gradually increases as the discharge proceeds[20]. Therefore, it is necessary to clarify how the conductivity variation affects the initiation and development of discharge inside the bubble. Figure 6 shows the ICCD images of the first discharge inside a bubble in solutions with different conductivities. As the solution conductivity increases, both the number and the luminous area of the discharge channels inside the bubble increase noticeably, and the discharge tends to extend along the inner bubble surface. At a conductivity of 4.6 μS/cm, the discharge is mainly confined near the H.V. electrode and exhibits a corona-like mode. When the conductivity is increased to 16.3 μS/cm, the discharge changes into a streamer-like mode, and multiple streamer channels propagate in different directions toward the inner bubble surface. Such surface-propagating streamer discharges are favorable for many water-treatment applications, because the reactive species generated by the discharge can interact more directly with the surrounding liquid, thereby enhancing the concentration of dissolved active species. As the solution conductivity is further increased to 126.2 μS/cm, both the number and the brightness of the streamer channels increase further. When the solution conductivity reaches 1 mS/cm, the very first pulse already produces an intense bright discharge, and the bubble ruptures almost instantaneously. In addition, when strong discharge persists inside the bubble for a long period, the electrode may suffer from corrosion under the combined effects of high temperature and electrochemical reactions, resulting in reduced durability[21][22]. This trend is also consistent with the current waveforms shown in Figure 6, where the discharge current amplitude of a single pulse increases with increasing solution conductivity. It should be noted that the measured current waveform includes both the displacement current and the discharge current. Therefore, the current peak reflects the combined contribution of capacitive charging and discharge conduction.

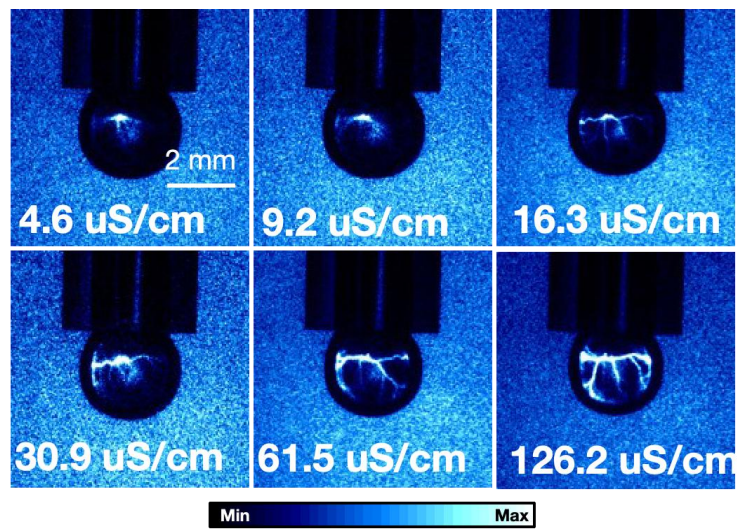

Figure 6. ICCD images of the first pulsed discharge inside a bubble in solutions with conductivities of 4.6, 9.2, 16.3, 30.9, 61.5, and 126.2 μS/cm. ICCD gate width: 900 ns.

The possible reasons for the propagation of streamer discharge along the inner surface of the bubble can be explained from the following two aspects. First, it is related to the spatial distribution of the electric field inside and around the bubble. Because the permittivity of the gas inside the bubble ($\varepsilon=1.0$) is much lower than that of the surrounding solution ($\varepsilon=78.0$), the electric field lines are refracted toward the gas–liquid interface, resulting in an enhanced local electric field in the interfacial region [23,24]. This field enhancement facilitates the initiation and propagation of streamer channels along the inner bubble surface. Second, since charges can accumulate on the inner bubble surface, the surface propagation of streamer discharge may be approximately regarded as a dielectric-barrier-discharge-like process occurring inside the bubble. The accumulated surface charges can modify the local electric field and thus guide the extension of the discharge along the bubble interface.

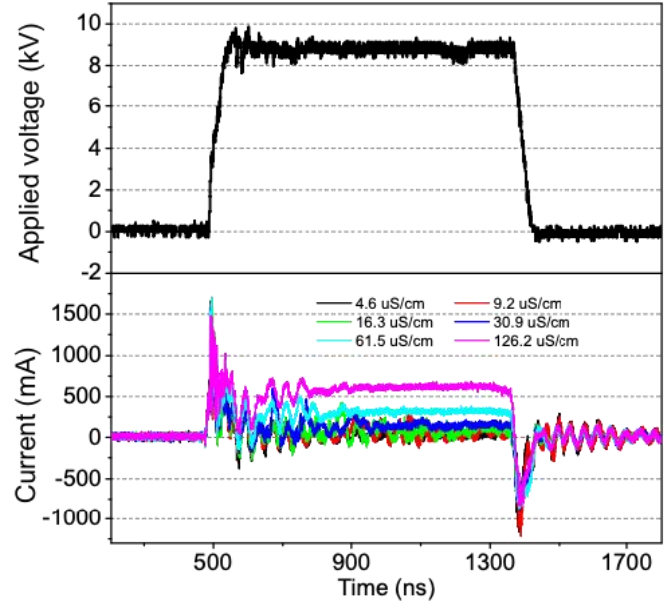

Figure 7 (a) Voltage and (b) current waveforms of the first bubble discharge in solutions with different conductivities.

Figure 8 shows the equivalent circuit of the underwater bubble discharge system, in which the bubble is modeled by an equivalent resistance $R_0$ and capacitance $C_0$, and the surrounding solution is modeled by an equivalent resistance $R_L$ and capacitance $C_L$. After a voltage pulse $V_0$ is applied, the charging process and the voltage distribution in the system are governed by the electrical impedances of the bubble and the liquid. At low solution conductivity, the liquid resistance $R_L$ is relatively high, so a larger voltage drop occurs in the liquid phase and the effective voltage across the bubble is relatively low, resulting in weak corona-like discharge. As the conductivity increases, $R_L$ decreases, allowing more voltage to be applied to the bubble region and promoting the transition from corona discharge to streamer discharge. The streamer channels then become brighter, more numerous, and more widely distributed along the inner bubble surface. At sufficiently high conductivity, the intensified discharge may propagate from the bubble interior into the surrounding liquid, causing rapid bubble rupture.

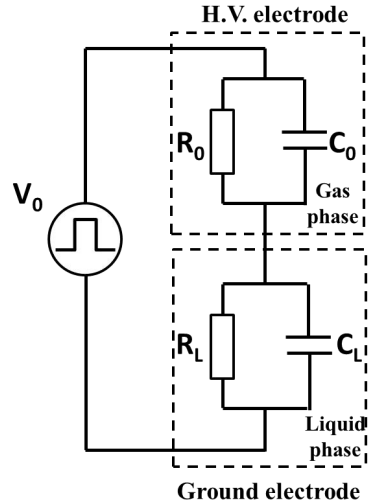

Figure 8. Equivalent circuit of preformed bubble reactor

In gas–liquid discharge applications, energy consumption is a key parameter for evaluating the practical applicability of the technology. The energy dissipated in a single pulsed discharge can be obtained by integrating the product of the voltage and current over time:

$$\varepsilon(t)=\int_0^t U(t)\times I(t)\ dt$$

where $\varepsilon(t)$ is the energy consumed by a single pulsed discharge, t is time, and U(t) and I(t) are the voltage and current, respectively. Figure 9 shows the relationship between the energy consumption per pulse of a single discharge inside the bubble and both the pulse number and the solution conductivity, with a pulse width of 900 ns. As shown in Figure 9 the energy consumption per pulse increases with both pulse number and solution conductivity. Under the same solution conductivity, the increase in pulse number leads to an increase in the average energy dissipated in a single pulse. This trend may be attributed to the accumulation of residual seed charges inside the bubble as the number of pulses increases, which enhances the discharge intensity and consequently raises the energy consumption per pulse. When the pulse number is the 1st and the solution conductivity is 126.2 μS/cm, the average energy consumption per pulse is about 3.9 mJ. When the pulse number increases to the 70th, the average energy consumption per pulse rises to about 4.8 mJ. In addition, for all pulse numbers, the energy consumption shows an approximately monotonic increase with conductivity, indicating that higher conductivity promotes stronger discharge development and greater energy dissipation during the pulsed discharge process.

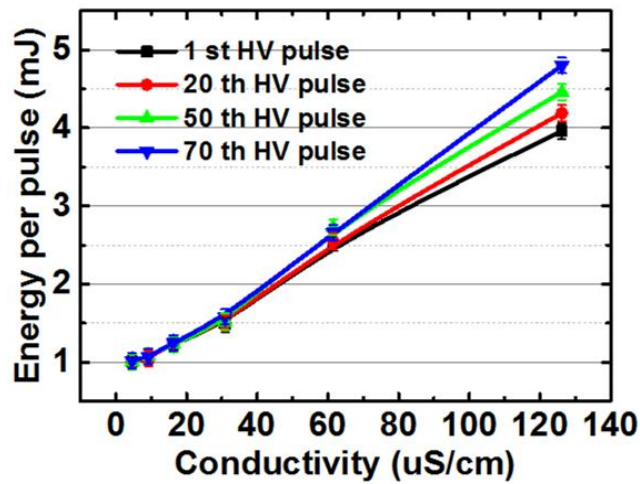

Figure 9. Energy per pulse as a function of solution conductivity and pulse number.

## 4. Conclusions

In this work, the evolution of pulsed discharge inside a preformed air bubble in water from the first to subsequent pulses was investigated experimentally. The main conclusions are as follows:

The synchronized needle–bubble system provided reproducible initial bubble conditions for investigating pulse-to-pulse discharge evolution inside a preformed bubble. Under identical experimental conditions, the first-pulse discharge inside the bubble was mainly corona-like and exhibited strong spatial randomness in both inception location and propagation path. The discharge probability increased with applied voltage and pulse number, indicating that residual seed charges and metastable species generated by previous pulses play an important role in facilitating subsequent breakdown.

The discharge behavior inside the bubble evolved significantly with pulse width, pulse number, and solution conductivity. With increasing pulse width and pulse number, the discharge gradually transitioned from a weak corona-like mode to a stronger streamer mode, accompanied by enhanced interfacial instability, bubble-surface wrinkling, and eventual bubble rupture under sufficiently strong excitation. Increasing solution conductivity promoted streamer initiation and propagation along the inner bubble surface, increased the discharge current amplitude, and raised the energy dissipated per pulse.

The present results demonstrate that the continuous application of voltage pulses is one of the key factors leading to discharge instability in repetitive bubble discharges. As the pulses proceed, residual effects and cumulative energy deposition continuously modify the discharge environment inside and around the bubble, thereby driving the transition from relatively stable early discharges to more unstable streamer development and interface disruption. These findings suggest that, in gas–liquid plasma reactors, control of pulse parameters and liquid conductivity is essential for improving discharge stability.

# Acknowledgments

This work was supported by the Liaoning Provincial Department of Education (Grant No. LJ212410165037).

.